\documentclass{webofc}
\usepackage[varg]{txfonts}   
\begin{document}
\title{A machine learning pipeline for autonomous numerical analytic continuation of Dyson-Schwinger equations}
%
%

\author{\firstname{Andreas} \lastname{Windisch}\inst{1,2,4}\fnsep\thanks{\email{windisch@physics.wustl.edu}} \and
        \firstname{Thomas} \lastname{Gallien}\inst{3,4}\fnsep\thanks{\email{thomas.gallien@silicon-austria.com}} \and
        \firstname{Christopher} \lastname{Schwarzlm\"uller}\inst{3,4}\fnsep\thanks{\email{christopher.schwarzlmueller@silicon-austria.com}}
}

\institute{Department of Physics, Washington University in St. Louis, MO 63130, USA 
\and
           Know-Center GmbH, Inffeldgasse 13, 8010 Graz, Austria
\and
           Silicon Austria Labs GmbH, Inffeldgasse 33, 8010 Graz, Austria
\and
	   RL Community, AI AUSTRIA, Wollzeile 24/12, 1010 Vienna, Austria
          }

\abstract{%
	Dyson-Schwinger equations (DSEs) are a non-perturbative way to express $n$-point functions in quantum field theory. Working in Euclidean space and in Landau gauge, for example, one can study the quark propagator Dyson-Schwinger equation in the real and complex domain, given that a suitable and tractable truncation has been found. When aiming for solving these equations in the complex domain, that is, for complex external momenta, one has to deform the integration contour of the radial component in the complex plane of the loop momentum expressed in hyper-spherical coordinates. This has to be done in order to avoid poles and branch cuts in the integrand of the self-energy loop. Since the nature of Dyson-Schwinger equations is such, that they have to be solved in a self-consistent way, one cannot analyze the analytic properties of the integrand after every iteration step, as this would not be feasible. In these proceedings, we suggest a machine learning pipeline based on deep learning (DL) approaches to computer vision (CV), as well as deep reinforcement learning (DRL), that could solve this problem autonomously by detecting poles and branch cuts in the numerical integrand after every iteration step and by suggesting suitable integration contour deformations that avoid these obstructions. We sketch out a proof of principle for both of these tasks, that is, the pole and branch cut detection, as well as the contour deformation.   
}
\maketitle
\section{Introduction}
\label{intro}
Computing Green's functions in the complex domain numerically turns out to be a tedious task, even when working in Euclidean space and when considering perturbative calculations at 1-loop level. The main complications one has to deal with are accounting for the poles and branch cuts that arise in the complex plane of the radial component of the square of the loop momentum, when the latter is expressed in hyper-spherical coordinates. Since the loop momentum is a 4-momentum, and because there is an external momentum flowing into the diagram, as well as the internal (loop-) momentum, two of the hyper-spherical angles can be integrated trivially, and one has to consider a two-fold, non-trivial integration over the square of the loop momentum, and over (the cosine of) the angle between the external and the internal momentum. As discussed e.g. in \cite{Windisch:2013dxa}, branch cuts are induced by the angular integration and in conjunction with the current value of the external and internal momentum square, and they appear in the complex plane of the radial integration variable. Because these branch cuts often intersect the positive real axis, the original integration contour of the radial part, that starts at the origin (or a small infrared (IR) cut-off) and runs along the positive real half-axis to some ultraviolet (UV) cut-off, has to be deformed around the obstructive branch cut to connect the origin and the UV cut-off, and without picking up the residue of any pole in the plane that could appear between the deformed contour and the line between the origin and the UV cut-off. Put into other words, the integration contour has to be continuously deformable from the original contour to the new, deformed contour that avoids the branch cuts.
While for perturbative loop diagrams usually a full, explicit form of the integrand is known, this normally does not hold true for DSEs (see e.g. \cite{Alkofer:2000wg} for a thorough treatment of the DSE formalism), as these are of the second Fredholm type and require e.g. iterative strategies to find a numerical solution. The dressing functions that result from a successful numerical solution of a Dyson-Schwinger equation appear on both sides of the equation, that is, explicitly on the left hand side, and in the integrand of the self-energy loop on the right hand side of the equation. The dressing functions are usually initialized to some real value, and given this as an initial condition, one could in principle analyze the integrand to find the poles and cuts for this given situation. One could then, in principle, find suitable contour deformations and avoid the cuts and poles, perform one iteration step, analyze the resulting, updated dressing functions, and try to deduce the newly arisen analytic obstructions for the next iteration step. Then, new contour deformations could be sought, and another iteration step could be performed. However, this is not feasible because of mainly two reasons. One, once we allow the dressing functions to take complex values for the external momentum square, the complex momenta will produce different branch cuts for each of its values, which results in a plethora of differently parametrized integration contours that have to be found. Usually one can cover all necessary contours by introducing a few parametrizations for different regions of the complex plane, but it is still a tedious task. The second reason why this is not feasible in practice is simply the fact that the system's solution is sought by an iterative approach, that is, after every iteration step such a full analysis would have to be conducted. 

In these proceedings we propose a possible way to overcome these obstacles by introducing a machine learning pipeline that analyzes the current analytical properties of the integrand, and finds suitably deformed integration contours autonomously. In more concrete terms, here and in the following, we provide evidence for the feasibility of such a pipeline by providing proof (or proof of concept) for each and every of the following steps:
\begin{enumerate}
\item \textbf{PROOF (A):} Suitability of numerical contour deformation to compute the analytic properties of Green's functions
\item \textbf{PROOF OF CONCEPT (B):} Feasibility of determining the analytic properties of numerically computed functions by means of deep learning and computer vision
\item \textbf{PROOF OF CONCEPT (C):} Feasibility of autonomous contour deformation by means of a deep reinforcement learning agent
\end{enumerate}
Here, the difference between \textbf{PROOF} and \textbf{PROOF OF CONCEPT} is, that the former has been successfully applied in a broad variety of contexts, which can thus be regarded as a successful strategy (and not as a proof in the mathematical sense). The latter means, that these steps have been introduced with the very application discussed here in mind, and we have published (and yet unpublished) knowledge that strongly suggests that the respective strategies will be successful.
These proceedings are organized as follows. In the following three sections we discuss the points \textbf{(A)}, \textbf{(B)}, and \textbf{(C)} in detail, and then conclude with a summary and outlook.

\section{Contour deformation for numerical computation of Green's functions in the complex domain \textbf{(A)}}
\label{sec-1}
In order to establish numerical proof that such a deformation can be obtained, we look at this study \cite{Windisch:2012zd}, in which a correlator of so-called $i$-particles was considered. These $i$-particles are discussed in \cite{Baulieu:2009ha}, where also an exact solution of the analytic properties of the correlator, that has been treated numerically in \cite{Windisch:2012zd}, is presented. The exact solutions of \cite{Baulieu:2009ha}, and the numerical results found in \cite{Windisch:2012zd} by means of contour deformation, are in perfect agreement. While \cite{Windisch:2012zd} is concerned with showing agreement between the analytic and numerical approaches, there is an abundance of studies in which this strategy has been deployed, see e.g. \cite{Fischer:2020xnb,Eichmann:2019dts,Miramontes:2019mco,Eichmann:2021vnj,Windisch:2012sz} for an exemplary but incomplete list of such studies.  

\section{Pole and branch cut detection with deep learning \textbf{(B)}}
\label{sec-2}
In this section we will discuss the pole and branch cut detection in numerical data by means of deep learning. In principle, and at least as far as pole detection is concerned, one could also do this by more conventional means, such as computing contours around possible pole locations in order to check for a non-vanishing residue, see e.g. \cite{Windisch:2016iud}. This doesn't resolve the detection of branch cuts, though, which is why we chose a different approach here. Using a so-called U-Net, that has been originally developed for biomedical image segmentation \cite{ronneberger2015unet}, we created a similar pipeline for detecting poles and branch cuts in snippets of numerically computed functions. The U-Net architecture is comprised of a convolutional neural network (CNN) \cite{Fukushima1980} with a contracting and an expanding path. In between, feature maps are being cropped and copied from the contracting to the expanding path. While images (in our case numerically computed snippets of functions in the complex plane) are being fed into the network, at its other end a segmentation mask is being produced, which, again in our case, marks the position of the poles and branch cuts. The network is trained by comparing the segmentation mask produced by the model with the ground truth that is provided as a 'label' for this learning setup.  

\subsection{U-Net based segmentation}
\label{ssec-21}
In order to generate a data set used for training the U-Net for pole and branch-cut segmentation, we again considered the problem of the $i$-particle correlator mentioned above \cite{Baulieu:2009ha}. The numerical data has been produced by solving the angular integral for complex internal momenta $(y)$ and complex external momenta $(x)$ in equation (22) of \cite{Windisch:2012zd} for $x,y\in[-10,10]\times i[-10,10]$. Once the cut has been produced, we added a random number of poles (between 0 and 9), with random positions and random residues additively to the function with the cut. The segmentation mask is a matrix with the same size as the data with the cut and poles ($128\times128$ pixels), with its entries being ternary numbers. Using base-3 covers the three possible cases of every positional value, namely the 'pixel' being a pole location, belonging to a branch cut, or being a point that is neither of these two cases. The pole locations can be directly entered in the segmentation mask once the random location has been produced for a given training sample. The branch cut position can be marked by using the cut parametrization equation (25) in \cite{Windisch:2012zd}.

\begin{figure}[h]
\centering
\includegraphics[width=6cm,clip]{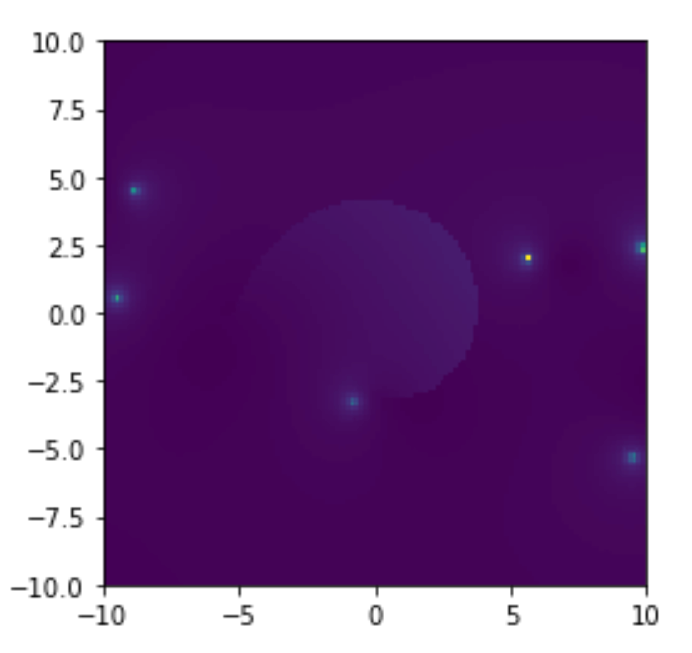}
\includegraphics[width=6cm,clip]{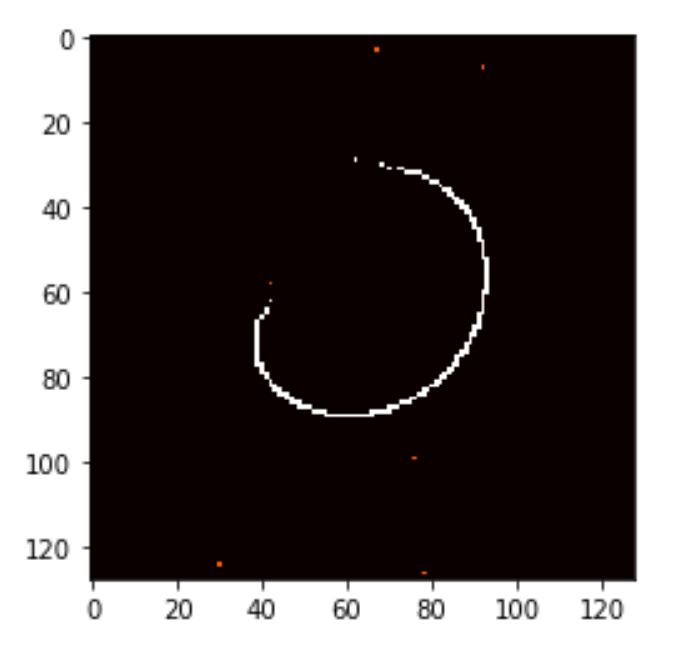}
\caption{Left: A training sample with a branch cut and poles, produced as described in the text. The horizontal and vertical axis correspond to the real and imaginary part of the loop momentum square, $y$, expressed in GeV$^2$. Right: The corresponding segmentation mask. Note that the segmentation mask appears rotated here, because it has been expressed in pixel space rather than in the physical coordinate system.}
\label{fig-1}       
\end{figure}

In Figure \ref{fig-1} we show an example of a training sample used to train the U-Net. 

\begin{figure}[h]
\centering
\includegraphics[width=10cm,clip]{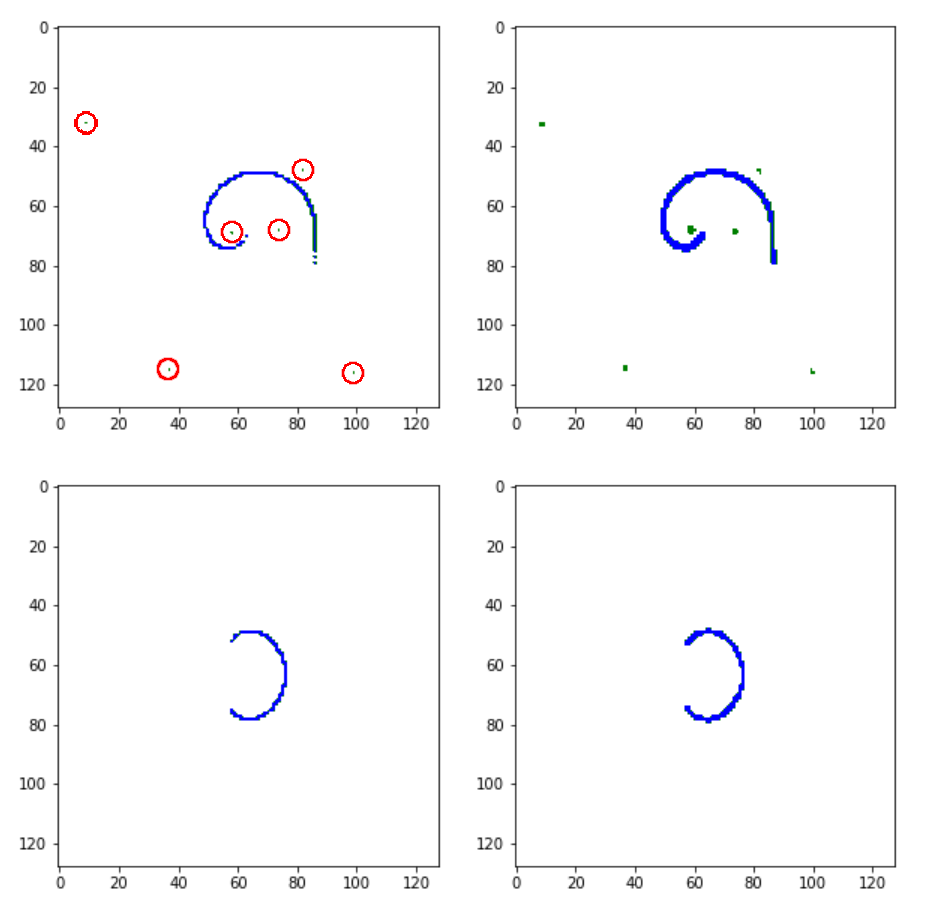}
\caption{Preliminary results with the first U-Net attempt. Left column: Ground truth of segmentation masks. Since each pole only occupies a pixel in the mask, they have been marked by small circles for readability. Right column: Predictions produced by the trained U-Net.}
\label{fig-2}       
\end{figure}

In Figure \ref{fig-2} we show preliminary results of the first, trained U-Net using a training set generated as described above. We are currently investigating various extensions to our approach and will thus not provide any hyper parameters of the model at this point, as we are preparing a separate publication for the pole and cut detection with deep learning at the moment. The results of this first attempt look very promising, and suggest that this approach can be used in a Dyson-Schwinger setup to detect the cuts and poles in the integrand automatically. There is, of course, the problem of generalizability that is still to be addressed. In order to employ this approach to a Dyson-Schwinger setup, the variability encountered in the course of the iterative procedure must be captured by the distribution of the data set the U-Net has been trained with, otherwise it will most likely fail to recognize the respective patterns. This problem mostly affects the branch cuts, as poles are point-like and do not possess any relevant sub-structure, while cuts are essentially 1-dimensional objects. One possible way to generate cuts of arbitrary shape could be to produce a finite-length cut using a logarithm, and then apply an arbitrary distortion to the resulting cut-line to  allow for arbitrary shapes.
Another issue that is relevant in the context of the approach as described here is the sparsity of the ternary masks. Since there are only a few poles, and since also the branch cuts do not occupy a large amount of pixels, there is a class imbalance between pixels labeled as poles, cuts, or neither of the two. This skewness can become an issue when computing the loss function when training the system. To this end, one could compress the labels by deploying a compressive sensing approach, as discussed in \cite{Xue:2017aa,Xue:2019aa} in the context of biomedical imaging. We are currently also investigating this option, but do not have results yet.

\section{Automated contour deformation through deep reinforcement learning \textbf{(C)}}
\label{sec-3}
In this section we discuss the next step in the pipeline, where we assume that the previous step has been successfully conducted and a segmentation mask that contains the information about the poles and cuts for the situation under consideration has been produced. Given this information, we must now find a suitable integration contour that avoids the poles (i.e. is continuously deformable from the line along $\mathbb{R}_+\cup\{0\}$), as well as the branch cuts. To this end we trained a deep reinforcement learning agent to perform this task (see \cite{Windisch:2019byg}). Deep reinforcement learning is a learning approach inspired by behavioral science, in which a learning entity (the agent) observes and lives in an environment, conducts actions in this environment, and receives a reward or punishment by means of a scalar signal (see e.g. \cite{DRL,Sutton:2018aa} for an overview and a text book on the subject). The desired behavior of the agent can then be learned by maximizing the expected, cumulative reward, and the trained agent is then in possession of a policy, that takes in the state of the environment, and produces the most suitable action for a given situation.
\subsection{Deep Reinforcement Learning agent for contour deformation}
\label{ssec-31}
\begin{figure}[h]
\centering
\includegraphics[width=8cm,clip]{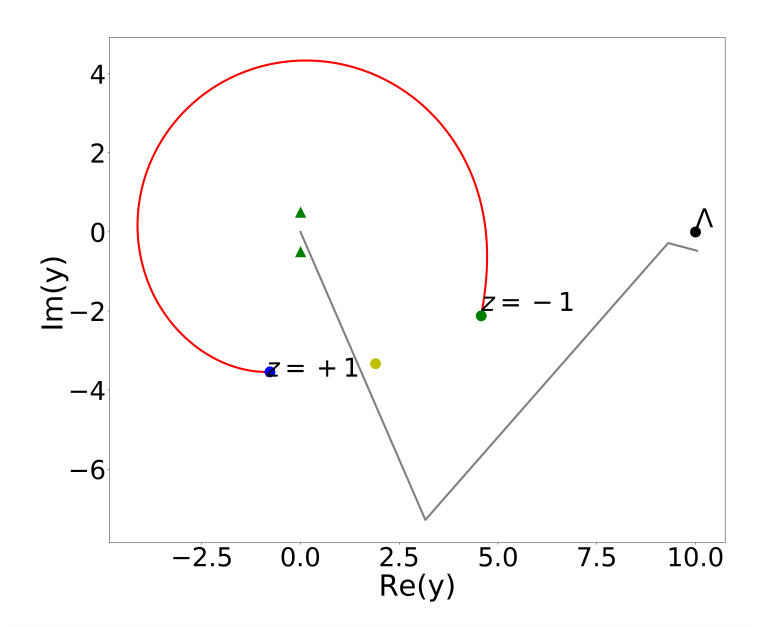}
\caption{A successful contour found by the agent published in \cite{Windisch:2019byg}. The red line represents the branch cut, the green triangles the complex conjugate poles located at $y=\pm\frac{i}{2}$. Note that this contour is continuously deformable from the line along the positive real half-axis to the cutoff, here represented by the $\Lambda$. The yellow point that is passed closely by the contour represents the direction of the opening of the cut, an information made available to the agent.} 
\label{fig-3}       
\end{figure}

In \cite{Windisch:2019byg}, we considered a continuous scenario, in which we again used the $i$-particle correlator of \cite{Baulieu:2009ha}. This correlator possesses a branch cut, as well as two complex conjugate poles. Using a proximal policy optimization based approach \cite{Schulman:2017aa}, we successfully trained an agent that could produce meaningful contours in this environment, see Figure \ref{fig-3}. The environment in this case consists of the complex plane of the radial integration variable, and the agent perceives this environment by means of a vector that holds relevant information, such as where the cut is, where the poles are located, in which direction the cut opens up, etc. In principle, one could also train a deep reinforcement learning agent solely on the numerical data as input, however, we decided against this approach, as we would lose a substantial amount of control over the whole setup, which, in addition, would be much more involved as well.
In order to deploy this approach for DSEs, the DRL agent will be re-implemented in a discrete setup, as the continuous environment is harder to solve and not necessary for the relatively small sizes of discrete points the correlator will be solved for.

\section{Conclusions and outlook}
\label{sec-4}
In the previous sections we provided proof, as well as proof of concept, for the main ingredients needed to set up an autonomous machine learning pipeline to perceive the analytic properties of the numerically computed, radial integration plane of the self energy loop in a DSE, as well as to produce the corresponding deformed integration contours. The first step, identifying the non-analyticities, will be implemented using either a U-Net architecture, or a regular CNN with dense labels produced by a compressive sensing approach. Once this information is available, it will be prepared for the DRL agent to observe, which will then produce a suitable contour. This could be done in every iteration step of the DSE, thereby solving the equation, while respecting the constraints as imposed by the non-analyticities. In order to study the quark propagator Dyson-Schwinger equation in Landau gauge in Euclidean space in the complex domain, we plan to deploy a suitable truncation scheme that establishes minimal constraints through the analytic properties of the self-energy loop.  

\section*{Acknowledgements}
AW acknowledges support through the U.S. Department of Energy, Office of Science, Office of Nuclear Physics under Award Number \#DE-FG-02-05ER41375.

%
%
%

%

\bibliography{citations}

\begin{thebibliography}{18}

\bibitem{Windisch:2013dxa}
A.~Windisch, M.Q. Huber, R.~Alkofer, Acta Phys. Polon. Supp. \textbf{6}, 887
  (2013), \texttt{1304.3642}

\bibitem{Alkofer:2000wg}
R.~Alkofer, L.~von Smekal, Phys. Rept. \textbf{353}, 281 (2001),
  \texttt{hep-ph/0007355}

\bibitem{Windisch:2012zd}
A.~Windisch, R.~Alkofer, G.~Haase, M.~Liebmann, Comput. Phys. Commun.
  \textbf{184}, 109 (2013), \texttt{1205.0752}

\bibitem{Baulieu:2009ha}
L.~Baulieu, D.~Dudal, M.S. Guimaraes, M.Q. Huber, S.P. Sorella,
  N.~Vandersickel, D.~Zwanziger, Phys. Rev. \textbf{D82}, 025021 (2010),
  \texttt{0912.5153}

\bibitem{Fischer:2020xnb}
C.S. Fischer, M.Q. Huber, Phys. Rev. D \textbf{102}, 094005 (2020),
  \texttt{2007.11505}

\bibitem{Eichmann:2019dts}
G.~Eichmann, P.~Duarte, M.T. Peña, A.~Stadler, Phys. Rev. \textbf{D100},
  094001 (2019), \texttt{1907.05402}

\bibitem{Miramontes:2019mco}
A.S. Miramontes, H.~Sanchis-Alepuz, Eur. Phys. J. \textbf{A55}, 170 (2019),
  \texttt{1906.06227}

\bibitem{Eichmann:2021vnj}
G.~Eichmann, E.~Ferreira, A.~Stadler (2021), \texttt{2112.04858}

\bibitem{Windisch:2012sz}
A.~Windisch, M.Q. Huber, R.~Alkofer, Phys. Rev. \textbf{D87}, 065005 (2013),
  \texttt{1212.2175}

\bibitem{Windisch:2016iud}
A.~Windisch, Phys. Rev. \textbf{C95}, 045204 (2017), \texttt{1612.06002}

\bibitem{ronneberger2015unet}
O.~Ronneberger, P.~Fischer, T.~Brox, \emph{U-net: Convolutional networks for
  biomedical image segmentation} (2015), \texttt{1505.04597}

\bibitem{Fukushima1980}
K.~Fukushima, Biological Cybernetics \textbf{36}, 193 (1980)

\bibitem{Xue:2017aa}
Y.~Xue, N.~Ray, CoRR \textbf{abs/1708.03307} (2017), \texttt{1708.03307}

\bibitem{Xue:2019aa}
Y.~Xue, G.~Bigras, J.~Hugh, N.~Ray, IEEE Transactions on Medical Imaging
  \textbf{38}, 2632 (2019)

\bibitem{Windisch:2019byg}
A.~Windisch, T.~Gallien, C.~Schwarzlm\"uller, Phys. Rev. E \textbf{101}, 033305
  (2020), \texttt{1912.12322}

\bibitem{DRL}
V.~Fran{\c{c}}ois{-}Lavet, P.~Henderson, R.~Islam, M.G. Bellemare, J.~Pineau,
  CoRR \textbf{abs/1811.12560} (2018), \texttt{1811.12560}

\bibitem{Sutton:2018aa}
R.S. Sutton, A.G. Barto, \emph{Reinforcement Learning: An Introduction} (MIT
  Press, 2018)

\bibitem{Schulman:2017aa}
J.~Schulman, F.~Wolski, P.~Dhariwal, A.~Radford, O.~Klimov, CoRR
  \textbf{abs/1707.06347} (2017), \texttt{1707.06347}

\end{thebibliography}

\end{document}